\documentclass[letterpaper,fleqn,10pt,5p]{elsarticle}

\usepackage[numbers]{natbib}
\usepackage[T1]{fontenc}

\begin{document}


\title{A glint in the eye: photographic plate archive searches for non-terrestrial artefacts}

\author[1,2]{Beatriz Villarroel}
\ead{beatriz.villarroel@su.se}
\author[1]{Lars Mattsson}
\author[3]{Hichem Guergouri}
\author[4,5]{Enrique Solano}
\author[6,2]{Stefan Geier}
\author[7]{Onyeuwaoma Nnaemeka Dom}
\author[8]{Martin J. Ward}

\address[1]{Nordita, KTH Royal Institute of Technology and Stockholm University, Roslagstullsbacken 23, SE-106 91 Stockholm, Sweden}
\address[2]{Instituto de Astrof\'isica de Canarias, Avda V\'ia L\'actea S/N, La Laguna, E-38205, Tenerife, Spain}
\address[3]{Science of the Matter Division. Research Unit in Scientific Mediation, Constantine, Algeria}
\address[4]{Departamento de Astrof\'isica, Centro de Astrobiolog\'ia (CSIC/INTA), PO Box 78, E-28691, Villanueva de la Ca\~{n}ada, Spain}
\address[5]{Spanish Virtual Observatory}
\address[6]{Gran Telescopio Canarias (GRANTECAN), Cuesta de San Jos\'{e} s/n, 38712 Bre\~{n}a Baja, La Palma, Spain}
\address[7]{Center for Basic Space Science, National Space Research and Development Agency, Enugu-Nigeria.}
\address[8]{Centre for Extragalactic Astronomy, Department of Physics, Durham University, South Road, Durham, DH1 3LE, UK}

\begin{abstract}
In this paper, we present a simple strategy to identify Non-Terrestrial artefacts \citep[NTAs;][]{Kopparapu} in or near geosynchronous Earth orbits (GEOs). We show that even the small pieces of reflective debris in orbit around the Earth can be identified through searches for multiple transients in old photographic plate material exposed before the launch of first human satellite in 1957. In order to separate between possible false point-like sources on photographic plates from real reflections, we present calculations to quantify the associated probabilities of alignments. We show that in an image with nine ``simultaneous transients'' at least four or five point sources along a line within a $10 \ast 10$ arcmin$^{2}$ image box are a strong indicator of NTAs, corresponding to 
significance levels of $2.5$ to $3.9 \sigma$. This given methodology can then be applied to set an upper 
limit to the prevalence of NTAs with reflective surfaces in geosynchronous orbits.  
\end{abstract}


\maketitle

\section{Introduction} 
\label{sec:intro}
Searches for technosignatures have never been more popular and received more attention than in recent years. The Searches for Extraterrestrial Intelligence (SETI) have been carried out in the radio band since the 1960s \citep{Cocconi,Drake1960}, and since 2015, the Breakthrough Listen \citep[see e.g.][]{Lebofsky,Price} initiatives have deployed some of the world's powerful radio telescopes in the most extensive SETI effort undertaken to date. For example China's giant FAST radio telescope is conducting SETI observations. While most SETI searches have been carried out in the radio, there are reasons to believe that artefact searches and optical transient surveys may be more successful \citep{Shostak,Davenport}. Both targeted and untargeted searches for laser signatures have been reported.  \cite[e.g.][]{Tellis,Villarroel2020,Isaacson,Marcy2021}. Meanwhile, alternative new concepts for how to search for ET are emerging, \citep{Singam,Hector2021}. Many of these ideas propose searching for ET far away, in distant galaxies or around other star systems.

Of these different research topics within SETI, solar system SETI has received far less attention than others. This is paradoxical, as humans have demonstrated the motivation and the capacity to send exploratory probes to other stellar systems. For example, the \textit{Breakthrough Starshot} program plans to launch a probe to Alpha Centauri within twenty years. If humans have these capacities it seems natural to search for extraterrestrial probes or other so-called ``Non-Terrestrial Artifacts'' \citep[NTAs;][]{Bracewell,Kopparapu} inside our Solar System.

In this paper we present an unexplored and, in terms of its feasibility, straighforward SETI strategy. We attempt to answer the question whether a space-faring civilisation has undertaken surveillance of Earth by means of physical space probes. A space-faring civilisation in the distant past ($>$ 100,000 years ago) may have sent probes to explore the Earth and under these circumstances some of these probes would have remained in high-altitude orbits around Earth. Of particular interest are the geosynchronous Earth orbits (GEOs) currently populated by communications satellites. GEO satellites nearly always remain over same region on Earth. It is tenable that the GEOs might also be used by another civilisation to study the Earth. A probe or spacecraft no longer in use, could in principle survive for billions of years in GEO, before impact with meteorites and collisions disintegrate eventual probes into very small debris pieces of only a few centimeters size. It has been shown that radiation pressure \citep{HectorTalk} does not remove debris at GEO orbits, at least not on Myr timescales.

Satellites that are uniformly illuminated at low- or medium altitude orbits leave clear streaks in the long time exposures from old photographic plates as they move at speeds projected as hundreds of arcseconds per second. At higher or GEO altitudes the presence of satellites or space debris can be detected by fast, transient glints caused by surface reflection of the Sun. When the reflective surface of the satellite coincides perfectly with the position of the observer and the Sun, a short but powerful glint can be observed. Despite the fast movement of the satellite, the very brief reflective alignment means that the resulting short duration glint has a Point Source Function (PSF)-like shape -- indistinguishable from fast astrophysical transients e.g. afterglows from GRBs or FRBs. Indeed, searches for astrophysical transients show that most of the transients detected in the automated transient surveys today are just solar reflections from artificial objects in GEOs. These solar reflections usually occur on short time scales $t \sim 0.2$ or 0.4 seconds \citep{Nir2020,Corbett2020,Karpov2016} and with peak magnitudes ranging from $9 - 11$ mag, corresponding to fragment sizes of a few tens of centimeters. Sometimes, multiple transients are observed within a few minutes from each other within the same field of view \citep{Nir2020}. Multiple transients could be the result of either several debris fragments glinting as sunlight refects off their surfaces, or one single fragment glinting as it tumbles.

Searches for glints have revealed an important fact about space debris. We know now that the majority (75 \%) of glints from the GEO are not associated with any known object listed in the USSTRATCOM catalogue and must therefore be centimeter-sized space debris \citep{Blake2020}. This space debris will not dissipate rapidly and in time it increases further the risk of debris generation through cascades of collisions (the so-called Kessler syndrome). Environmental pollution of Space resulting from numerous satellite launches, leaves a longterm deleterious imprint on near Earth orbits that should not be ignored. This demonstrates the advantage to carry out specific searches for non-human artificial objects in the GEOs.

To search for these structures in modern surveys as Pan-STARRS \citep{Chambers,MagnierA,MagnierB,MagnierC,Waters2016,Flewelling2016} or Evryscope \citep{Law2016,Ratzloff2019} is a colossal challenge requiring fine modeling of tracks and surveys due to all the human contamination. Lacki (2019) \cite{Lacki2019} proposes a program capable of detecting glints with large ``spot size'' from artificial, spinning objects located further than the Moon, where each glint could last for hours, rather than seconds, and leaving a ``train'' of glints in the eyes of the observer. While, in principle this is correct, dedicated search efforts in modern sky surveys are costly and time-consuming.

In this article we explain why broad based programs such the Vanishing \& Appearing Sources during a Century of Observations \citep[VASCO;][]{Villarroel2020,VillarroelCSP} are extraordinarily well suited to detect signs of artificial structures in the GEOs or even low-Earth orbits while comparing old photographic plate material of our sky, taken before the first satellite was launched in 1957, with modern imaging. While the VASCO program aims to search for vanishing stars, its methodology is capable of discovering other, unexpected features in the data. By piggybacking on VASCO's ``vanishing star'' classifications, one has ample opportunity to collect glints from artificial objects.

In the next section we discuss possible origins of reflective artefacts in GEOs. We then discuss the lifetime of space debris in GEOs and suggest some specific signatures to be looked for using photographic plate material, including some examples of what has been found so far. In the final section (``Discussion'') we discuss how GEO probes and their resulting debris in GEO offer a direct, and mostly overlooked, technomarker in SETI and how programs such as the VASCO project \citep{Villarroel2020,VillarroelCSP} can be amongst the best opportunities to yield a first detection of artificial remnants  -- \textit{if they exist}.

\section{The origin of non-terrestrial artefacts}
\label{sec:origin}

The possibility of the presence of non-human technology, or NTAs, within our Solar System has been explored in a limited way and mostly at the theoretical level in the astronomical literature \citep{Gertz2016,Gertz2020}. 
Despite strong arguments in favour of searches for Solar System artefacts, so far there have been very few observing programs to test the whether or not the Solar System is devoid of NTAs \citep{Kopparapu}. In contrast, the prospect of finding NTAs has provoked much interest in popular culture and in science fiction literature.

In defining NTAs there are two generic kinds. First are objects that are operational or ``active'' which may appear as unaccountable transient events. The second category are objects that are non-operational or ``passive'' whose orbits could be manifested by the glint timing depending on their underlying geometry, precession and spin.

The first category includes exploratory probes from other civilisations. Since Earth's civilisation have proposed a plan to send probes to another stellar system through \textit{Breakthrough Starshot}, it is therefore reasonable to assume that other civilisation might also be motivated to explore our Solar System \citep{Bracewell}. Some of these probes could be in orbit around Earth or pass straight through the Solar System. Albeit continuing to be highly controversial \citep[see e.g.][]{Curran}, it has been proposed that the {\it 'Oumuamua} interstellar object is a space-craft \citep{Bialy}, based on a number of its anomalous characteristics. 

If they exist such probes could be in plain sight on the surface of the Moon, asteroids, trojans, minor planets or on Mars \citep{Kopparapu,Davies}. Such objects might reveal themselves through emissions in the radio or microwave or visible light \citep{Loeb2012}. The various Lagrangian points have been investigated for probes to some degree, but without finding any evidence \citep{Freitas,Valdes}. However, ``lurking'' probes might have been designed to be concealed in co-orbitals \citep{Benford2019,Benford2021}. A recent article  \citep{Hector2021} has proposed a possible mission employing ultra-high resolution imaging in combination with machine learning techniques, that could search for probes on other planets or astronomical bodies. An active effort in space archaeology has recently been initiated \textit{Galileo}  project\footnote{https://projects.iq.harvard.edu/galileo/activities} that looks for interstellar visitors and unexpected aerial objects within Earth's atmosphere.

The second category of passive objects are more challenging to explore. Consider for example, that long ago ie. tens of thousands or even millions of years, a space-faring civilisation had sent probes to the Earth. Having fulfilled its mission or exhausted its fuel, the advanced civilisation then lost contact with it,leaving only the inert remains of the probes behind. These probes would still exist in orbit around the Earth (or further out in the Solar System). They might follow low orbits, moderate-height orbits, geosynchronous ones or even reside in the  high \textit{graveyard orbits}, where our Space agencies have placed some redundant satellites. They could be single or in groups. If an object has a surface consisting of reflective material, this may occasionally glint. Even a small piece of metal that has become detached from the main machine, will glint. Some other materials with high albedo, e.g. certain polymers, also have the necessary reflective surface.

We now consider some other scenarios. For example could previous civilisations have existed on Venus, Mars or even the Earth itself, a so-called {\it prior indigenous technological species} \citep{Wright2017,Schmidt2018}? Even if there were to be some technological signatures one could expect a previous civilisation on the Earth to leave on the surface of the planet, plate tectonics would obliterate geological evidence for all artifacts on a timescale of million years. However, according to those who support the controversial concept of the Anthropocene, some evidence for any advanced long past technologically advanced civilisation on Earth might still persist despite plate tectonics. For example if they launched objects into orbit. For Venus, it was long assumed that liquid water once existed on the surface of the planet for about 3 billion of years and during its habitable period complex life might have emerged \citep{Way2020}, until around 700 million years ago. This possibility has been challenged since it is now known that there is a very low density of water in Venus' atmosphere \citep{Hallsworth2021}. Mars on the other hand, is known to have once had lakes and rivers of liquid water some billions of years ago, when it had a much thicker atmosphere that could retain water molecules. Also all the building blocks necessary to create complex organic molecules are present on the planet. But so far there is no evidence that primitive life, that might have evolved into an advanced civilisation, ever existed on Mars. However, as for the Earth, eventually plate tectonics would remove evidence of previous civilisations. The advantage of searching for NTAs of previous civilisations in space itself, is that there is no plate tectonics in space. We could potentially see left-overs from previous civilisations -- provided that the civilisation did not exist too long time ago (several billion years ago) as the orbits are likely to have been cleared from artefacts during such a long time period.

 \begin{figure*}\label{Hichem1}
  \resizebox{\hsize}{!}{
  \includegraphics{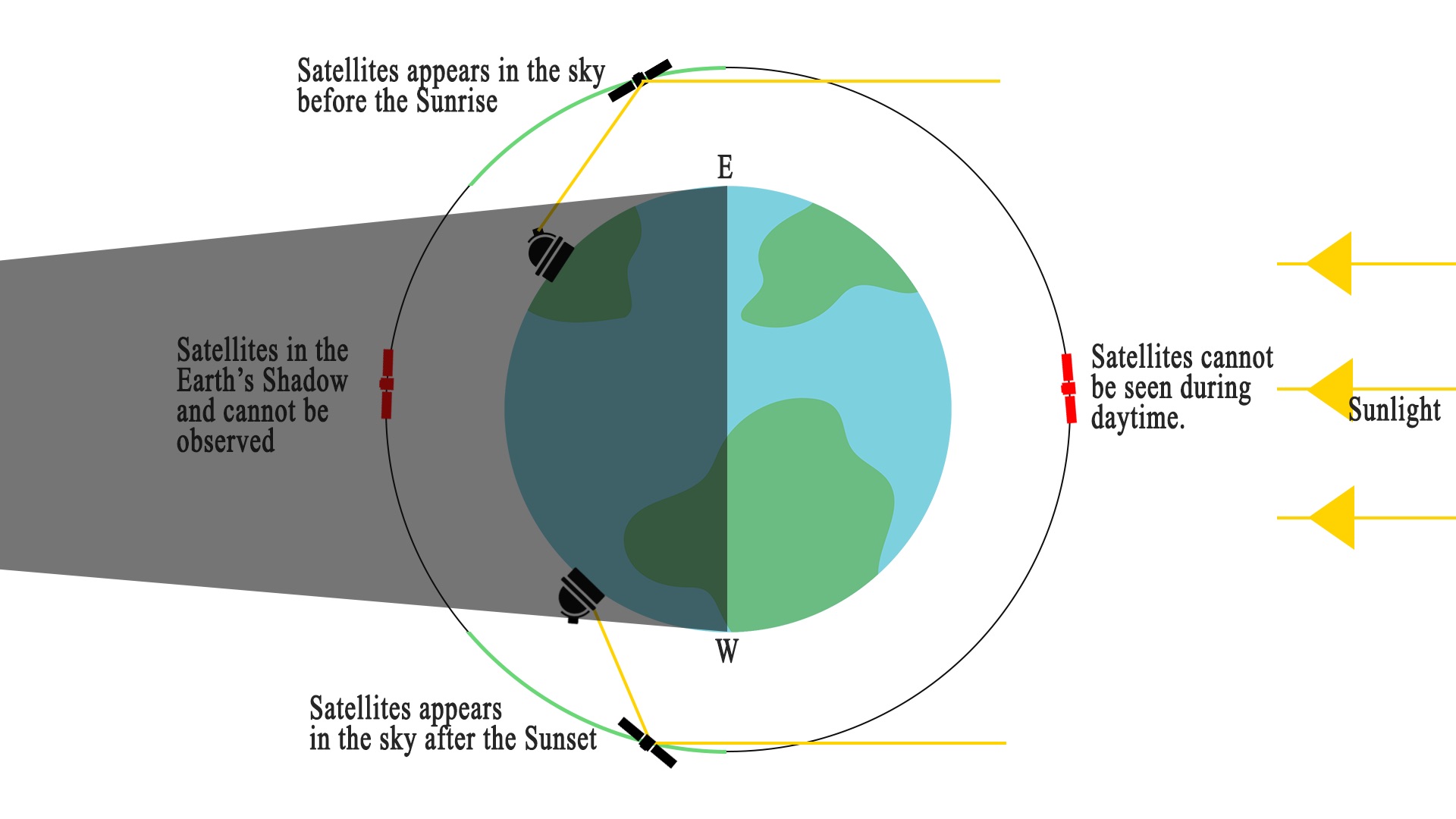}}
  \caption{\textbf{Illumination of satellites.} Satellites are only visible when the Sun illuminates them and the background sky is not too bright. The higher the altitude of the orbit, the longer the satellite is outside the shadow of the Earth.} 
   \end{figure*}

\section{Life-time of debris in space}
\label{sec:lifetime}
The total amount of space debris in orbit around Earth is now enormous. It is estimated using statistical models that about 36,500 objects larger than 10 cm are in orbit; the number of small objects ranging between 1 to 10 cm in diameter is approximately 900,000, and there are 128 million (!) particles that are even smaller \cite{ESA}. Among all this space debris evidence for a non-terrestrial artefact might easily be missed.

Assuming that another civilisation has left NTAs in geosynchronous orbits, it is useful to know for long could reflective pieces of e.g. metal or glass (or other reflective materials) remain in orbit? The determining factors are (1) gravity: debris at too low altitude, will fall back to the Earth within a few years. (2) collisions with natural objects will break up the objects into many smaller pieces, and also eject some out of orbit, and (3) radiation pressure from the Sun. Wright (2017) \cite{Wright2017} argues that these natural factors will result in the artefacts surviving for less than a few Gyr in orbit. Due to gravitational perturbations their orbits will also drift over long time scales.

The question of how long objects can stay in a geosynchronous orbit around the Earth was discussed by Socas-Navarro \textit{(TechnoClimes 2020}) \cite{HectorTalk}. They sought to estimate how long a ``Clarke exobelt'' around another star could be visible from Earth. A Clarke exobelt is a dense belt of debris at a geosynchronous orbit that might be detected via the transit light curves of exoplanets, so providing a potential technomarker \citep{Hector}. The author estimated that at the currently much accelerated rate of satellite launches, within 200 years the Earth will then have a detectable exobelt. He also estimated that these satellites can survive in orbit for Myrs, making it a durable technomarker. These satellites will be subjected to collisions among each other, that will cause the pieces to break into smaller fragments, causing a wider spread over the belt. Some objects might have high density, and others lower, which means they will have different ballistic coefficients, and will not be equally prone to stay in orbit. Another consequence of collisions is that the total cross-section of all fragments is larger than that of a single object, so making it more likely to detect both a ``glint'' from Earth, or a Clarke exobelt around another planet. Socas-Navarro \textit{(TechnoClimes 2020}) \cite{HectorTalk} concludes that the survival of objects in a GEO is at minimum $\sim 10^{5}-10^{7}$ years if only considering the collisions, and likely to be on a $\sim$ Gyrs time scale.

A different problem to consider is how long a fragment of flat metal or glass can stay in space and remain reflective, given collisions with dust grains, micrometeorites and possible radiation damage. Even very small micrometeorites $10 - 100$ $\mu m$ in size will cause much damage, especially as they hit the surface of a probe between $4 - 50$ km per second. These micro holes will eventually damage the flatness and reflectivity of the material. As an example, the rate of micrometeorites hitting the Moon has been estimated by studying lunar rocks as well as studying cubes made out of Suprasil-fused silica that were left on the Moon for 40 years. After this time about 10$^{-4}$ of the surface had been covered with micrometeorite hits \citep{Johnson1992}. This means that to cover the entire surface with micrometeorite hits, needs $\sim$100 000 years. For a material to survive longer than tens of thousands of years, it would require a self-repairing mechanism.

More challenges for objects in geosynchronous orbits arise due to their exposure to sunlight and energetic particles such as protons and MeV to GeV-cosmic rays. The material will eventually be damaged and lose its reflectivity. Studies of the effect of protons on mirror reflectivity \citep{Garoli2020} show how an aluminium mirror protected by a layer of SiO$_{2}$ loses its reflectivity when irradiated with protons at 60 keV $-$ 100 keV and an integral fluence of 10$^{17}$ protons cm$^{-2}$. The effect was strongest at short wavelengths, meaning the least affected is reflectivity in the red and infrared part of the spectrum. Using the typical integral fluence of protons at geosynchronous orbits of 10$^{14}$ protons per cm$^{-2}$ year $^{-1}$  \citep{Inguimbert2012}, we estimate that such a mirror would lose half of its reflectivity at optical wavelength within $\sim$1000 years.

In order to deflect protons and high energy particles, an extraterrestrial probe might have shielded material using a strong magnetic field during its cruise phase to the Earth and during its operational time, and so extend its durability. Since it would take about 73 000 years for a probe like Voyager to reach the nearest star Proxima Centauri, it is reasonable to assume that an extraterrestrial civilisation that launches a probe to the Earth will have developed materials and systems that could endure space travel of up to millions of years.

The degrading of material due to micrometeorites and cosmic radiation opens up a window of new possibilities: were one to make a mission to the geosynchronous orbits to collect the debris, it is almost trivial to identify debris that has been there for thousands of years by looking for objects having the most micrometeorite hits and largest loss of reflectivity of its surface.

\section{Signatures imprinted in photographic plates}
\label{sec:predictions}
The subject of this section is how to recognize signs of artificial objects in the pre-satellite images. For low orbits the evidence could be satellite streaks. Fig. 1 shows how the illumination of satellites is made visible from Earth. Typical low-orbit satellites that can be seen by the naked-eye do not emit light themselves, and are visible through the reflection of sunlight. Obviously they cannot be observed during daytime because of the sky brightness, and at night they are in Earth's shadow for most of the night. That means they are best observed either before sunrise or after sunset. They are seen as fast-moving point sources, giving rise to \textit{long, continuous streaks} in photographic images of long exposure. A satellite whose reflective surface is spinning gives rise to \textit{}{dashed} or broken lines. Continuous streaks may often be confused with natural objects e.g. meteors or asteroids. Closer inspection can be used to distinguish between meteors and satellites, for example \textit{fainting edges}.

Another signature is that if one has images of the 
entire sky, one knows that the satellites that are bound to Earth \textit{follow an arch}, while meteors go along straight lines. Asteroids often tumble, and moreover, can be identified through the JPL Horizon list of known asteroids. Also they have fainting edges in their streaks, as seen in Figure 2.

 \begin{figure*}\label{Asteroid}
 \resizebox{\hsize}{!}{
  \includegraphics{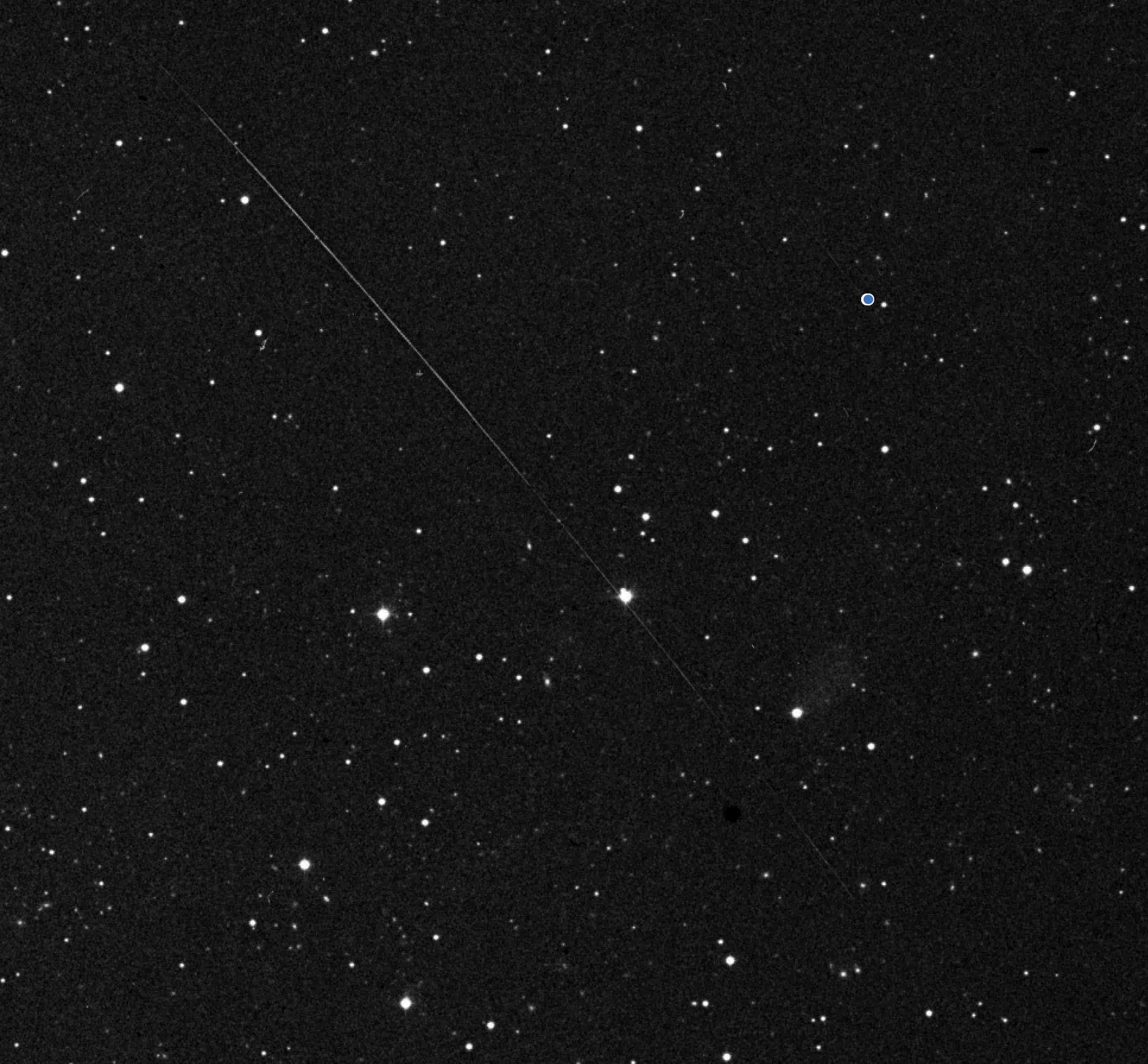}}
  \caption{\textbf{A typical streak.}} The POSS-I streak found in a red image identified through the citizen science project shows the effect of tumbling and is a possible near-Earth asteroid but is also a possible candidate. The streak is roughly 40 arcminutes in length and unlikely to be a meteor with its angular velocity and pattern of first being dim, then brightens, and then dims once again. The typical exposure time for POSS-I E images if about 50 minutes.
   \end{figure*}

For geosynchronous orbits what may be observed is a glint, when the Sun, the observer and the reflective surface are in the correct alignment. This is provided the satellite is not located within the Earth's shadow, where the sunlight will not make it glint (although, satellite lasers could). At higher altitudes the shadow cone of the Earth is much smaller. Two years ago the glint rate of human satellites \citep{McDowell} observed from Earth was 340 glints $h^{-1} sky^{-1}$ with $V < 4$ mag or 740 $h^{-1} sky^{-1}$ in dark locations with a sensitivity reaching $V \sim 6$ mag, where $h$ is hour and $sky$ means the whole sky. Near the equator about 1800 glints $h^{-1} sky^{-1}$ are predicted as many satellites follow geostationary orbits and cluster near to the equator.

From a single glint on an astronomical image it is not possible to determine whether it is a genuine short-lived transient or a satellite glint, based on just the PSF. For example, a recent paper \citep{Jiang2020B} reported an extremely unusual GRB flash apparently from the highest redshift galaxy at $z \sim 11$ \citep{Jiang2020A}. 
Shortly after its publication several other groups demonstrated that the observation was not a real transient, but in fact a satellite glint, see e.g. Steinhardt et al. (2021) \cite{Steinhardt}. Therefore what we require for our search is an indicator that cannot be confused with any natural phenomenon and is unlikely to be an instrumental defect. We will consider the signatures by which satellite glints manifest themselves within the observations. Beyond the presence of a single glint, the VASCO citizen science project could discover artificial objects in other ways:

\begin{itemize}
    \item \textit{Multiple glints with point source PSFs.} In particular these should be detected in the Kodak 103-aE emulsions (red plates) that is more sensitive to reflected sunlight (Fig. 3). A streak may appear to contain embedded glints, caused by a variation in brightness that is shorter than the exposure time.
    \item \textit{Glints along a line.} Multiple glints with typical PSFs that lie along a straight line, cannot be caused by any natural object or by any known type of plate defect (Fig. 4). They can arise when one single object, or a fragment of an object, on a particular orbit reflects sunlight as it spins. These glints can be, although they are not required to be, equidistantly placed along the line.
    \item \textit{Triple glint.} In one observation described by Deil et al. (2009)\citep{Deil2009}, a triple glint was observed using a Cherenkov type telescope. The two time intervals between the glints were identical, corresponding to a rotation frequency of $\sim$15 Hz. In this way a triple glint could be an observational signature of a rotating fragement of space debris. An example of a triple glint is shown in POSS-I, Fig. 5.
\end{itemize}

\begin{figure*}\label{fig:multy}
   \resizebox{\hsize}{!}{
   \includegraphics{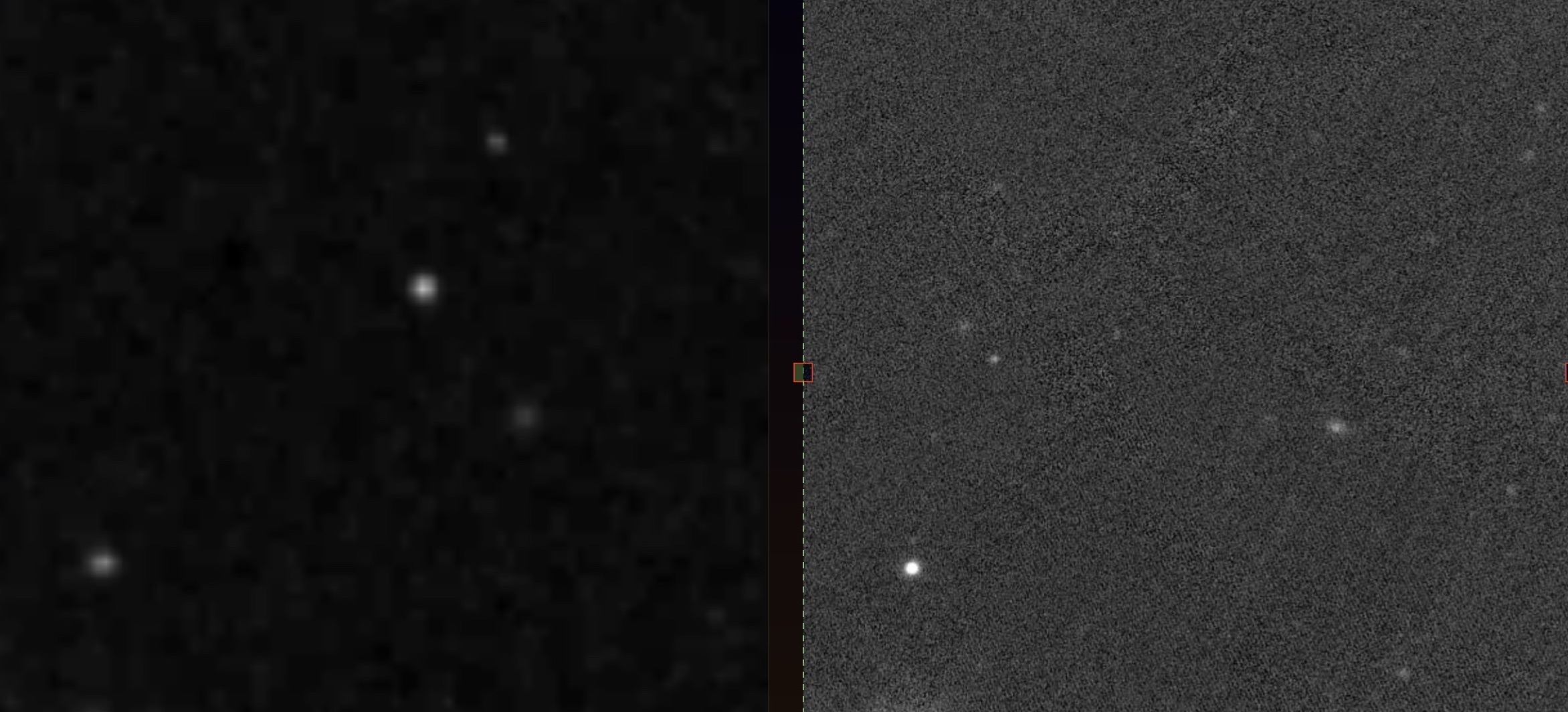}}
  \caption{\textbf{Multiple glints.} An example of multiple glints in a red POSS-I image from 1950s. The left column shows the POSS-I image, and the right column the Pan-STARRS image ($>$ year 2015). This example is based on the candidate from Villarroel et al. (2021)} \cite{Villarroel2021} and uses the VASCO citizen science web interface.
   \end{figure*}

  \begin{figure*}\label{fig:Euny}
  \resizebox{\hsize}{!}{
  \includegraphics{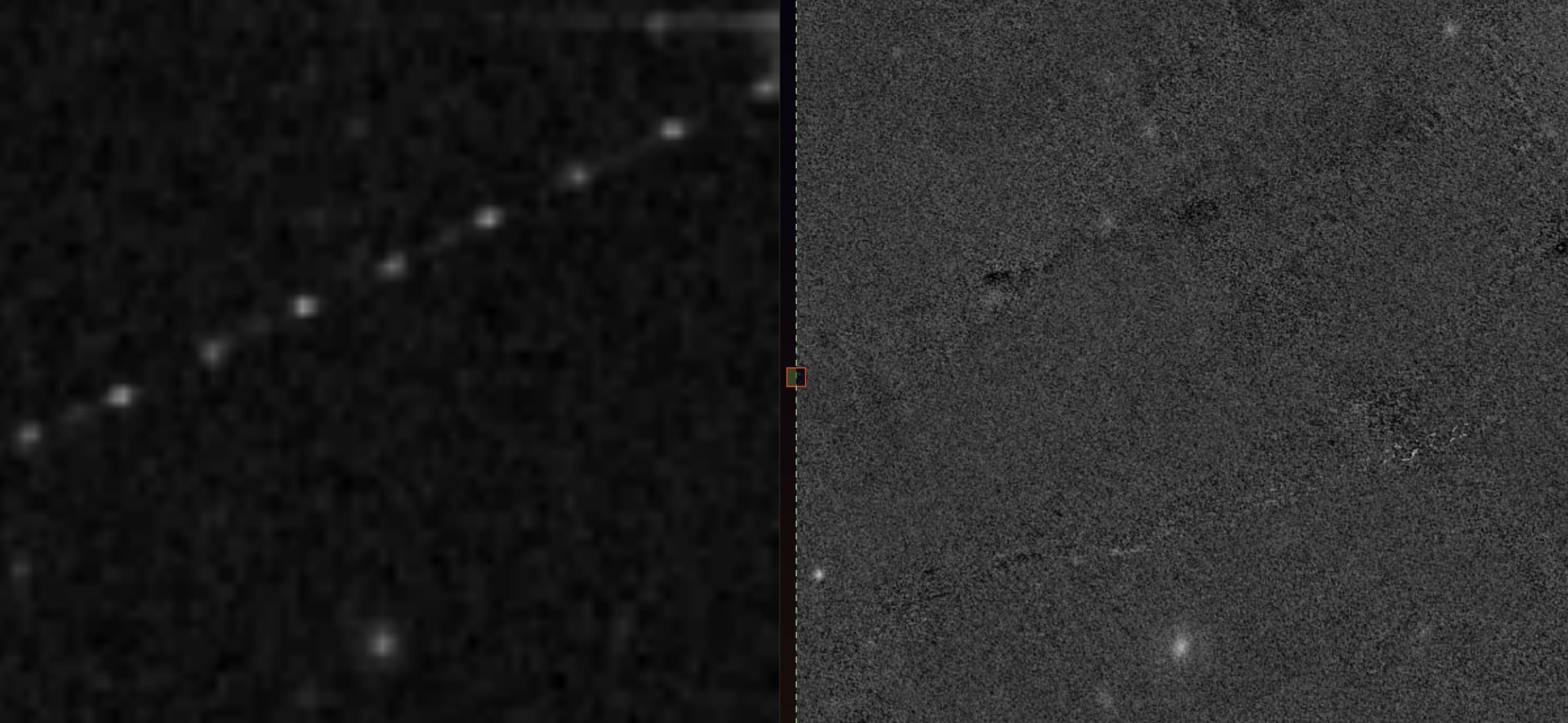}}
  \caption{\textbf{Repeating glints.} An example of repeating glints in a blue POSS-II image taken in 1980s,  here with a faint streak linking them together. The left column shows the POSS-II image, and the right column the Pan-STARRS image ($>$ year 2015). The example uses the VASCO citizen science web interface. An actual case of a ``train of glints'', can have significantly sparser spacing and can be composed of fewer than shown here.
   }
   \end{figure*}
   
\begin{figure*}\label{fig:Triple}
\resizebox{\hsize}{!}{
  \includegraphics{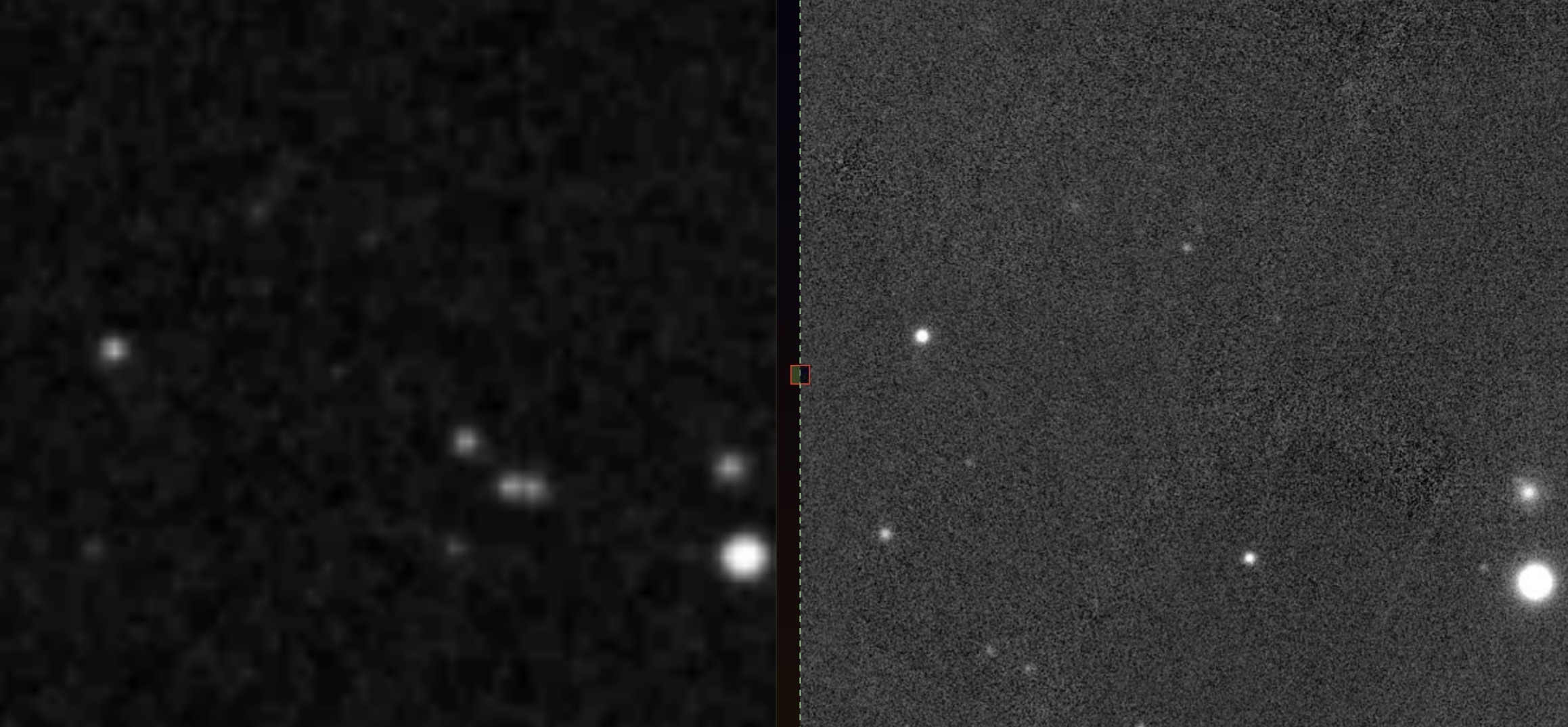}}
  \caption{\textbf{Triple glints.} An example of a triplet glints in a red POSS-I image from 1950s. The left column shows the POSS-I image, and the right column the Pan-STARRS image ($>$ year 2015). The example is from Villarroel et al. (2021) \cite{Villarroel2021} and uses the VASCO citizen science web interface.}
   \end{figure*}
   
 Clearly any object moving relative to the geostationary point, will leave a trail. These trails may appear like streaks, either continuous or with brightness variations depending on the shape of the object.
When the variation in brightness is longer than the exposure time, a line of varying thickness will result. However, these streaks can easily be confused with an asteroid track.

The signature of multiple glints is relevant as such an ``impossible’’ event was recently reported in a study by Villarroel et al. (2021) in Scientific Reports \citep{Villarroel2021}. The authors discovered nine simultaneous transients visible in a 10 by 10 arcmin$^{2}$ size image of the sky exposed in 1950. The objects were not visible half an hour earlier, or six days later. Follow-up observations with the 10.4-meter Gran Telescopio Canarias telescope could not link any counterparts to the original transient objects. Based on the time scales and the density of events, the authors ruled out all known astrophysical phenomena such as optical afterglows from gamma ray bursts, microlensing events, asteroids, meteors, variable or flaring stars. Known instrumental issues were also either excluded or deemed highly improbable. The authors concluded that if the detections are real, then the objects must be located inside the Solar System given their synchronous behaviour. It is premature to claim that such simultaneous transients represent non-terrestrial artefacts even if they were observed some years before Sputnik I was launched. For instance, contamination or emulsion defects could  coincidentally create false star-like imprints on old photographic plates, only discernible under investigation with a microscope. We therefore require another independent indicator. 

The \textit{smoking-gun observation} that settles the question unequivocally, is the one of repeating glints 
with clear PSFs along a straight line in a long-exposure image. When an object spins fast around itself and when its reflective surface faces the Earth, 
some of its parts could reflect sunlight. That results in multiple glints following a trail in an image. The number of glints might depend on the geometry 
and the speed of the rotation of the object. An object with only one single reflective surface that spins slowly will produce fewer glints 
than an object with several reflective surfaces that, moreover, spins fast. From the period one can also determine the shape of the glinting object.

An exciting aspect of these suggestions is that precisely these type of objects could be found during the course of the VASCO project \citep{Villarroel2020,VillarroelCSP}. Among the many objects classified as ``Vanished'', we could discover both single and multiple glint objects. Also through automatized methods, we seek to identify all cases of multiple glints within a small area of $10 \times 10$ arcmin$^{2}$, and to see if any of these represent cases where the glints follow a straight line.

The estimated number of ``multiple transients'', real or false, expected in, e.g., the POSS-I image data was estimated to about $\sim 0.07$ $h^{-1} sky^{-1}$ \cite{Villarroel2021}. That is negligible in comparison to the current number of satellite glints that would be detected by a telescope located near the equator sees today, $\sim$ 1800 $h^{-1} sky^{-1}$ \citep{McDowell}. Of these we expect that less than half will show at least three transients following a straight line.

It is possible that the method may lead to some cases being missed. For example an object coated in dark low albedo material, or an object that has been subjected to micro impact
and radiation over a very long time while in orbit, may be significantly less reflective. Also objects having only one small area reflective surface, might only display a single glint.

\section{Estimated probabilities}
\label{sec:probabilities}
To quantitatively estimate the likelihood that multiple transients which lie along a line represent satellite glints, we need to estimate the expectation value for the number of such alignments that would occur by just by chance. This can be done by investigating  $r$-point alignments, where $r$ is the number of transients along a straight line within a given field.

The simultaneous transients presented by \citet{Villarroel2021} appear to have three 3-point alignments in a cluster of just nine transients located in a $10\times 10$ arcmin$^2$ region. The probability of 3-point alignments occuring by chance in this situation is quite high. Based on \citet{Edmunds81,Edmunds85} we find that the expected number of 3-point alignments in a population of $N$ points (transients) defined by a strip of length $d$ and width $2\,p_{\rm max}$ located within an area $A$ on a plate is
\begin{equation}
    \mu = \frac{2\pi}{3}\frac{p_{\rm max}\,d_{\rm max}} {A^2}\,N\,(N-1)(N-2),
\end{equation}
where $d_{\rm max}$ is the maximal length of the strip/alignment, so that $d\le d_{\rm max}$.
If $N = 9$, $A = 100$~arcmin$^2$ (a $10'\times 10'$ square domain), $d_{\rm max}=10$~arcmin, $p_{\rm max} = 1.7$~arcsec (equal to image resolution) we get $\mu \approx 3$. If we assume $A$ is a circular-disc domain with a radius equal to the distance from the centre to a corner of the $10'\times 10'$ square domain, rather than $ A = 10'\times 10' = 100$ arcmin$^2$, the expectation value reduces to $\mu \approx 1$. Thus, we conclude that $\mu\sim 2$ is a reasonable estimate. In the image from \citet{Villarroel2021} the actual number of 3-point alignments is 3.

Clearly, we cannot argue that finding a 3-point alignment is interesting or significant. Not even two 3-point alignments in a sample of nine data points rises to the level of statistical significance. But if the number of objects in the alignment is 4 or more,then the situation changes drastically. A generalisation to an $r$-point alignment has the formula \citep{Edmunds85},
\begin{equation}
    \label{eq:rpoint}
    \mu = \frac{\pi\,2^{r-2}\,n^r\,p_{\rm max}^{r-2}\,A} {\Gamma(r-1)}\int_0^{d_{\rm max}} x^{r-1}\,e^{-2x\,n\,p_{\rm max}}\,dx,
\end{equation}
where $\Gamma$ is the gamma function and $n = N/A$.

For $2\,d_{\rm max}\,p_{\rm max}\,n \ll 1$ (which is applicable here for the most part) and integer $r$, this formula reduces to
\begin{equation}
    \mu \approx \frac{\pi\,2^{r-2}\,n^r\,p_{\rm max}^{r-2}\,d_{\rm max}^r\,A}{r\,(r-2)!}, \quad r = 3,4,5,...
\end{equation}
Using the example above, we see that $\mu \sim 10^{-2}$ if $r = 4$ and $\mu \sim 10^{-4}$ if $r = 5$ for the case of a circular-disc domain\footnote{The difference in $\mu$ values for the two cases with a square and circular-disc domains, respectively, is simply a factor $(\pi /2)^{r+1}$.}. Thus, a multiple transient event showing a 4-point alignment is perhaps sufficiently improbable to be a trustworthy indicator of glinting reflective material.

 Using the equation (\ref{eq:rpoint}) above we can also explore the number of expected $r$-point alignments from a more general point of view. If we express the equation for the special case when $2\,d_{\rm max}\,p_{\rm max}\,n \ll 1$ and $A = d_{\rm max}^2$,
\begin{equation}
    \mu \approx \frac{\pi\,N^r}{r\,\Gamma(r-1)}
    \left(\frac{2\,p_{\rm max}}{d_{\rm max} }\right)^{r-2},
\end{equation}
it is simple to  calculate $\mu$ as a smooth function of $N$ and $r$ for a given ratio of $p_{\rm max}$ and $d_{\rm max}$, provided that $r\ge 3$ and that we assume a plausible relationship between $A$ and $d_{\rm max}$, such as $A = d_{\rm max}^2$. 
In Fig. 6 we show expectation values obtained by numerical evaluation of equation (\ref{eq:rpoint}) for the number of $r$-point alignments within an area $A = d_{\rm max}^2$ containing $N$ points (transients), assuming $p_{\rm max}/ d_{\rm max} = 2.83\cdot 10^{-3}$ for consistency with the case considered above ($d_{\rm max}=10$~arcmin, $p_{\rm max} = 1.7$~arcsec). The blue zone corresponds to alignments that we expect to occur by chance in almost any set of points distributed in a plane. The red and yellow zones are more interesting as the expectation value is very low and such alignments, and 5 or more points are so extremely improbable to occur by chance that one of the most reasonable explanations is a series of glints from a near GEO object. Even in this extraordinary case, the finding must be followed up with more concrete searches for the probe as we do not know if other, currently inconceivable phenomena could give rise to a similar observational signature.

As we have already pointed out, even a 4-point alignment in an image with nine transients is an unlikely event, but a 1/100 probability ($\sim$ 2.5 $\sigma$) is not enough to completely rule out a random alignment. A five-point alignment is much better with a 1/10000 probability is much better ($\sim$ 3.9 $\sigma$). We note also that the number density of transients is a key parameter: already at $n = 20$ a 5-point alignment can easily happen by pure chance. Another important observation is that if just three transients show in an image, and these are aligned ($r = N = 3$), this is a quite improbable event compared to the expected frequency of such an alignment ($\mu \sim 0.1$). If we consider the case $r = N = 4$, we find $\mu \sim 10^{-3}$. Note, however, that all estimates made here are dependent on the ratio $\varpi_{\rm max} = p_{\rm max}/ d_{\rm max}$, which is a very uncertain parameter. For three-point alignments $\mu$ scales linearly with $\varpi_{\rm max}$, while in the case of four-point alignments ($r = 4$) the scaling is quadratic. Obviously, if $\varpi_{\rm max}$ is significantly larger than what we have assumed above, the number of expected random alignments goes up considerably. But $\varpi_{\rm max}$ is always a relatively small number, so $r \geq 4$ random alignments should always be rare events. Thus, looking for $r \geq 4$ events is a good strategy despite some evident uncertainties. We note, however, that a 3-point alignment could also be interesting and rare. For example a 3-point alignment found in a $10 \ast 10$ arcmin$^{2}$ image with in total three simultaneous transients, connected by a line no longer than 5 arcmin in length and 1 arcsec in width, would have a probability of 1 in 100. It is therefore always interesting to explore any r-point alignments.

  \begin{figure}\label{fig:mu_rpoint}
   \resizebox{\hsize}{!}{
  \includegraphics[trim=0.5cm 0.5cm 0.5cm 0.5cm, clip=true]{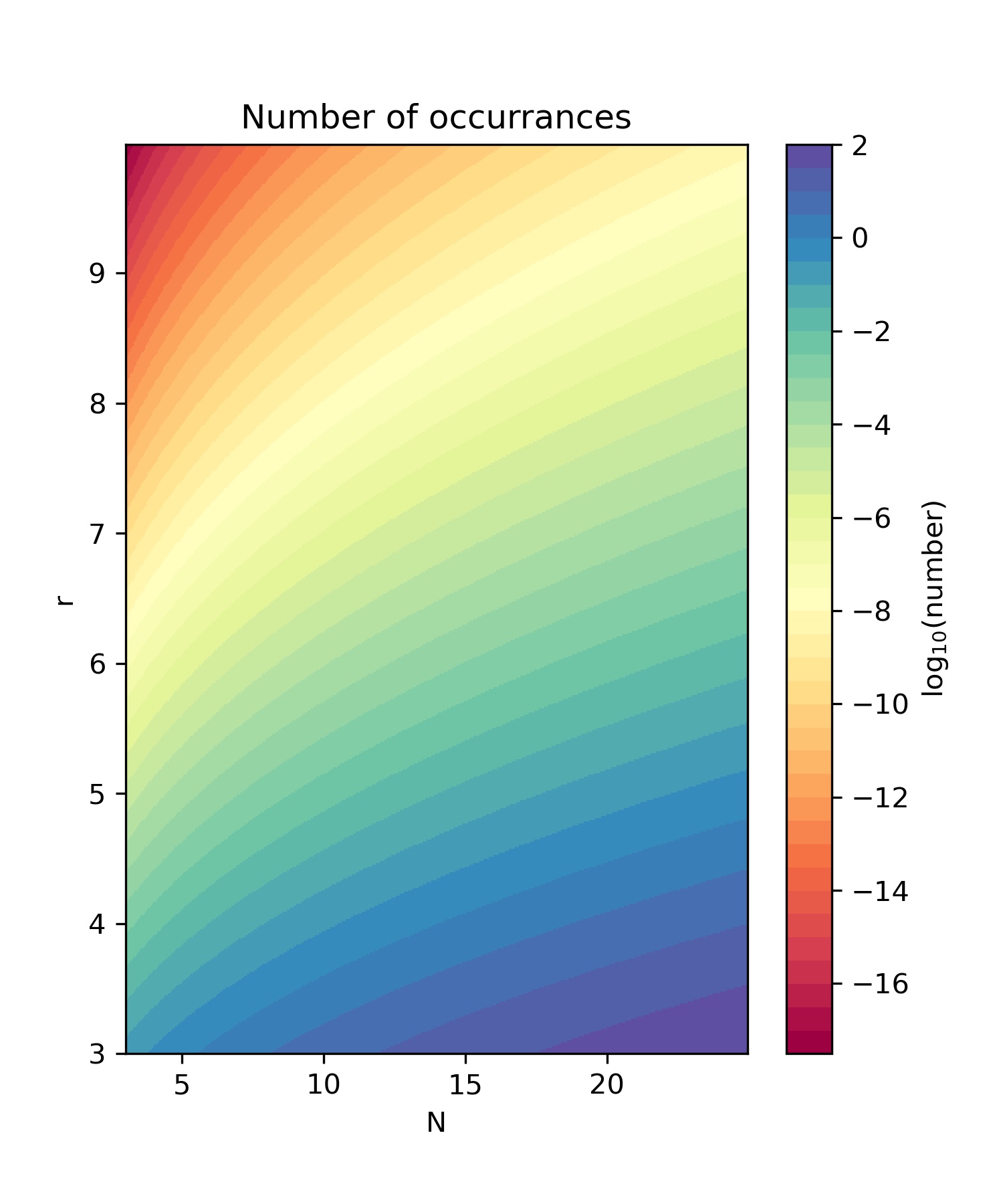}}
  \caption{\textbf{Expectation values.} Number of expected occurrences of $r$-point alignments within an area $A = d_{\rm max}^2$ containing $N$ points, assuming $p_{\rm max}/ d_{\rm max} = 2.83\cdot 10^{-3}$. Note that using a circular-disc domain, as discussed in the text, would change the expected occurrences s of $r$-point alignments by a factor $(\pi /2)^{r+1}$.} 
   \end{figure}

\section{Discussion}
\label{sec:summary}
In this paper we have discussed a variety of signatures of artificial objects near Earth and how they can be detected in pre-satellite imaging data. We demonstrate that a direct  signature are fast glints from reflective satellites or fragments thereof, in geosynchronous orbits. These glints occur when the observer, the Sun and the object are exactly aligned so that the observer sees the sunlight reflected off of the objects' reflective surface. Particularly bright and fast glints, occur when the objects are small, flat and mirror-like, e.g. artificial structures on satellites and space debris. A rocky surface such as that of an asteroid, has neither the shape or the necessary reflectivity to create the strong sub-second glints produced by an artificial object.

A glint resulting from a human satellite or other artefact space debris will often appear point source-like, with a PSF similar to that of a natural transient, even if the object is moving. This is because the glint is of short duration and the seeing of the observation has as FWHMs $>$ 1.5 arcsec. For example objects that track geosynchronous orbits at typical speeds of e.g. 15'' per second in images taken a seeing of $\sim$ 1.5 seconds, will look like a PSF if they glint faster than 0.1 seconds. In old photographic plates such as the First Palomar Sky Survey (POSS-1) plates, the seeing is sometimes significantly worse than is now typical of CCD-based observations.

In the intriguing case of the simultaneous transients described in \citep{Villarroel2021}, the seeing was very poor at $\sim 7$ arcseconds. A seeing of 7 arcseconds means that a glint from an object that travels 15''/second during the exposure time has to be shorter than $<$ 0.5 seconds in order to look like a PSF in the image. For the curious case of the simultaneous transients, that were found on long-exposure red emulsion images of 50 minutes, such a long exposure will necessarily dilute the observed flux. If these transients are real, the reported POSS-1 magnitudes of the simultaneous transients must be very much fainter than the actual apparent magnitude during the glint event. We can estimate how much, by using the exposure time of the POSS-1 plate -- about 50 minutes (or 3000 seconds) -- and assuming a 0.5 seconds duration glint from a geosynchronous satellite. This gives a flux dilution factor of $3000 s/0.5 s\sim 6000$, which corresponds to a reduction of about 9.4 magnitudes for the actual glint. We apply the correction to the simultaneous transients listed in Table 1 of Villarroel et al. (2021) \cite{Villarroel2021}. Only 5 out of 9 transients have their POSS-I magnitudes listed: three were not included as they did not have follow-up observations, and one was found in an overcrowded area. Correcting the magnitudes for the flux-dilution factor make them fall well within expectations of typical apparent magnitudes (about 8 - 10 mags) for glints arising from debris at the GEO \cite{Nir2020}. 

Provided these are the actual apparent magnitudes of the glints, the sizes of such possible objects must therefore be similar to the sizes of typical space debris fragments described in Nir et al. (2020) \cite{Nir2020}. Here, it is deduced that the physical objects are around a few tens of centimetres if the reflective surface is a type of transparent material, or even smaller of cm scale if it is perfectly reflective and mirror-like.
   
As a single occurrence, a case of simultaneously occurring transients in an image like the one in Villarroel et al. (2021) \cite{Villarroel2021} should not be taken as an evidence of satellites glinting at GEOs, due to that several randomly placed ``transients'' in an image might be the result of some unusual type of contamination or defects. However, their presence supports searches for other, clearer signatures of potential debris and satellites in orbits around Earth. The best way to search them, is obviously by looking at images taken before human-made objects were sent to orbit the GEOs.

This paper represents a motivation for the SETI community to use pre-satellite image data to engage in time domain searches for artificial objects in orbit around Earth, in particular highly reflective ones at high altitude. An object located in a geosynchronous orbit may have been there for many millions of years. Intact material or debris from degraded probes could easily be detected even if they have experienced multiple collisions during this time period. The recent \textit{Galileo} project is preparing to systematically search for these non-human, artificial structures in modern sky surveys over the coming years. However, pre-1957 archival photographic plates will be particularly powerful tools for this topic, since the sky was then free from human-made contamination.

A direct signature of solar reflections from artificial, reflective materials in geosynchronous orbits would be observation of multiple glints along a line in photographic plate images. Finding a single such case clearly merits careful on-site searches, necessary to obtain direct evidence. In view of the very rapid increase in satellite launches and human space debris in place at the GEOs, the time window for extremely short to assemble new datasets. Therefore we encourage SETI researchers to perform searches for glints in photographic plate material and to assist in VASCO's searches for these possible intriguing technosignatures. Currently VASCO's searches only target POSS-I data. The same investigations should be conducted using all photographic plate material, including digitized plates from the Lick and Sonneberg observatories and the \textit{Cartes du Ciel}.
\\

\section{Acknowledgments}
\label{sec:ackn}
The authors wish to thank Geoff Marcy, John Gertz, Ravi Kopparapu, Jacob Haqq-Misra and Hector Socas-Navarro for helpful and constructive comments on the manuscript. B.V. wishes to thank the Galileo project for providing a helpful and inspiring scientific community. B.V. is funded by the Swedish Research Council (Vetenskapsr\aa det, grant no. 2017-06372) and is also supported by the The L’Or\'{e}al - UNESCO For Women in Science Sweden Prize with support of the Young Academy of Sweden. She is also supported by M\"{a}rta och Erik Holmbergs donation.



\end{document}